\documentclass[runningheads,a4paper]{llncs}
\usepackage{url}
\usepackage{subfig}
\usepackage{epsfig}
\usepackage{multirow}
\usepackage{graphics}
\usepackage{graphicx}
\usepackage{algorithmic}
\usepackage{algorithm}
\usepackage{amsfonts}
\usepackage{amsmath}
\usepackage[parfill]{parskip}
\usepackage[applemac]{inputenc}

\usepackage{color}
\usepackage{array}
\usepackage{colortbl}

\newcommand{\bra}[1]{\langle #1|}
\newcommand{\ket}[1]{|#1\rangle}

\let\oldsqrt\sqrt
\def\sqrt{\mathpalette\DHLhksqrt}
\def\DHLhksqrt#1#2{%
\setbox0=\hbox{$#1\oldsqrt{#2\,}$}\dimen0=\ht0
\advance\dimen0-0.2\ht0
\setbox2=\hbox{\vrule height\ht0 depth -\dimen0}%
{\box0\lower0.4pt\box2}}

\begin{document}

\mainmatter  % start of an individual contribution

\title{Can Quantum Entanglement Detection Schemes Improve Search?}
\titlerunning{Can Quantum Entanglement Detection Schemes Improve Search?}

\author{Luís Tarrataca\inst{1}\thanks{Luís Tarrataca was supported by FCT (INESC-ID multiannual funding) through the PIDDAC Program funds and FCT grant DFRH - SFRH/BD/61846/2009.} \and Andreas Wichert\inst{1}}
\authorrunning{L. Tarrataca \and A. Wichert}

\institute{
GAIPS/INESC-ID\\
Department of Informatics\\
IST - Technical University of Lisbon - Portugal\\
\email{\{luis.tarrataca,andreas.wichert\}@ist.utl.pt}\\
}

%\date{\today}

\maketitle

\begin{abstract}

Quantum computation, in particular Grover's algorithm, has aroused a great deal of interest since it allows for a quadratic speedup to be obtained in search procedures. Classical search procedures for an $N$ element database require at most $O(N)$ time complexity. Grover's algorithm is able to find a solution with high probability in $O(\sqrt{N})$ time through an amplitude amplification scheme. In this work we draw elements from both classical and quantum computation to develop an alternative search proposal based on quantum entanglement detection schemes. In 2002, Horodecki and Ekert proposed an efficient method for direct detection of quantum entanglement. Our proposition to quantum search combines quantum entanglement detection alongside entanglement inducing operators. Grover's quantum search relies on measuring a quantum superposition after having applied a unitary evolution. We deviate from the standard method by focusing on fine-tuning a unitary operator in order to infer the solution with certainty. Our proposal sacrifices space for speed and depends on the mathematical properties of linear positive maps $\Lambda$ which have not been operationally characterized. Whether such a $\Lambda$ can be easily determined remains an open question.

\keywords{quantum computation, tree search, entanglement detection}

\end{abstract}

%\pacs{03.65.Ud, 03.67.Ac}

%\keywords{quantum computation, tree search, entanglement detection}

%\maketitle

\section{Introduction \label{sec:introduction}}

Computer scientists are often faced with the task of constructing algorithms capable of delivering a solution for a given problem. For some problems it is possible to engineer algorithms capable of producing a solution with a number of computational steps that is bounded by a polynomial $n^{k}$ where $n$ is the length of the input and $k$ some constant. The class of problems for which a polynomial-time algorithm exists is known as P. Problems belonging to P are usually seen as being efficiently solvable, i.e. tractable. Class EQP represents the quantum equivalent of P.

For other problems it is possible to verify in polynomial-time if a given configuration is a solution, although there are no known methods for efficiently calculating a solution. For these type of problems, there is no alternative but to perform an exhaustive search of all possible configurations. The class NP consists of those problems whose possible configurations can be verified in polynomial-time. Clearly, $\text{P} \subseteq \text{NP}$ since the possibility of constructing a solution in polynomial time also implies that a solution can be verified efficiently. One of the outstanding questions in computer science consists in determining if the class NP is equivalent to the class P, i.e. P=NP? Traditionally, approaches to answering this question have focused at a subclass of NP, namely NP-complete problems. This subclass contains those problems which are both NP and NP-hard. A problem is said to be NP-hard if an algorithm capable of solving it can be translated into an adequate algorithm for any NP problem. By its own definition, an efficient solution for a problem in NP-complete implies that an efficient solution exists for all problems in NP.

The first clues that some problems which are classically hard may have an efficient quantum solution were provided in \cite{deutsch1992}. Shor's algorithm for efficient factorization \cite{shor1994} reinforced this idea. Later, Grover's search algorithm \cite{grover1996} provided an asymptotical quadratic speedup over classical strategies. The quantum search algorithm systematically increases the probability of obtaining a solution with each iteration. After the algorithm has concluded, a measurement is performed in a quantum superposition, in order to obtain a solution with high probability. The superposition state represents the set of all possible results. Grover's approach sparked interest by the scientific community on whether it would be possible to devise a faster search algorithm. Unfortunately, it was proved that the search problem cannot be solved under $\Omega(\sqrt{N})$ time \cite{bennett1997} using standard quantum computation approaches. 

In this work we present an alternative search method based on the principles of tree search decomposition and quantum entanglement detection. Unlike traditional approaches, we opt not to concentrate our efforts on measuring a quantum superposition of possible values. Rather, we are more interested in exploiting the unitary operator that is applied to a quantum superposition in order to infer possible solutions with certainty. However, an implicit caveat exists associated to our quantum search proposal. Namely, our system implies a tradeoff between speed and space that will become apparent in the following sections.

The next sections are organized as follows: Section \ref{sec:traditionalApproachesToTacklingNP-CompleteProblems} focuses on presenting the details of an NP-complete problem, namely the Boolean satisfiability problem, alongside classical tree search techniques of examining the problem space. Section \ref{sec:approach} presents our hybrid approach, combining tree search decomposition alongside with quantum entanglement detection schemes. Section \ref{sec:conclusions} presents the conclusions of this work.

\section{Traditional approaches to tackling NP-Complete problems \label{sec:traditionalApproachesToTacklingNP-CompleteProblems}}

The satisfiability (SAT) problem was the first problem ever shown to be NP-complete \cite{cook1971}. SAT asks whether a given boolean formula is satisfiable. Any polynomial-time algorithm capable of solving SAT automatically enables an efficient solution for all of NP. In complexity theory, the satisfiability problem is a boolean formula $\phi$ composed of \cite{cormen2001}

\begin{itemize}

	\item $n$ boolean variables: $x_{1}, x_{2}, \cdots, x_{n}$;
	
	\item $m$ boolean connectives: any boolean function with one or two inputs and one output, such as $\land$ (AND), $\lor$ (OR), $\neg$ (NOT), $\rightarrow$ (implication) and $\leftrightarrow$(if and only if);
	
	\item parentheses.
	
\end{itemize}

We are interested in determining a set of values for the variables of $\phi$, i.e. variable configuration, which cause the overall expression to be satisfiable, i.e. evaluate to true. At any given point in time we need to consider the $n$ variables alongside $m$ gates, i.e. we can verify any configuration in $n + m$ time. However, the number of possible configurations to consider grows exponentially with the cardinality of the variable set. As an example lets consider the simple formula presented in Expression \ref{eq:satExample}.

\begin{equation}
\phi = ( x_{1} \land x_{2} ) \lor x_{3}
\label{eq:satExample}
\end{equation}

\begin{figure}[ht]
	\centering
	\includegraphics[width=10cm]{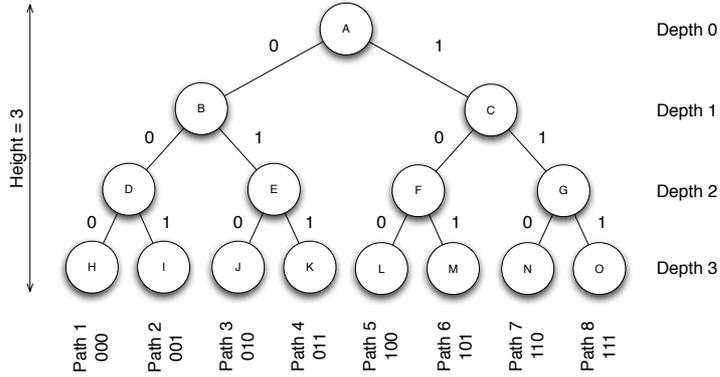}
	\caption{The possible paths for a binary search tree of depth $3$. \label{fig:binaryTreeExample}}
\end{figure}

The standard approach to solve such a problem would be to enumerate all possible configurations of the $m$ variables. This procedure can be better understood with the help of a simple tree diagram such as the one illustrated in Figure \ref{fig:binaryTreeExample}. At each depth level a specific the possible values for a specific binary variable are considered, e.g. depth $0$ considers the possible values for $x_{1}$, depth $1$ considers variable $x_{2}$ and so on. With each depth level an additional binary variable is taken into account. Considering $n$ binary variables requires examination of $2^{n}$ possible leaf states, i.e. $\Omega(2^{n})$. In tree search vocabulary these states are also known as paths. If the specific case of Expression \ref{eq:satExample} is mapped into the tree elements of Figure \ref{fig:binaryTreeExample} then paths $1$, $4$, $6$, $7$ and $8$ would evaluate to true.

\section{Approach \label{sec:approach}}

How can we proceed by developing an alternative approach to that of Grover's? First lets start by considering the following scenario: suppose we have a bipartite quantum system respectively labeled as the query register, $\ket{q}$, and the answer register, $\ket{a}$, acting on Hilbert space $\mathcal{H} = \mathcal{H}_{q} \otimes \mathcal{H}_{q}$. The query register is an $n$-qubit register where possible values for the binary variables of the SAT problem will be setup, i.e.  $\ket{q} = \ket{x_{1}x_{2} \cdots x_{n}}$. Notice that in order to gain a quantum advantage over classical computation we need to place $\ket{q}$ in a uniform superposition of the computation basis. This can be done efficiently by applying the Hadamard transform $H$ a total of $n$ times to the $n$-qubit state $\ket{0}$, i.e. $H^{\otimes n} \ket{0}^{\otimes n} =  \frac{1}{\sqrt{2^{n}}}\sum_{x = 0}^{2^{n} - 1} \ket{ x } $. Such a procedure enables the creation of a superposition containing an exponential number of states, each of which representing a possible tree path, by only employing a polynomial number of gates. The answer register contains a single qubit which is initialized to state $\ket{0}$. The overall state of the system can thus be described as illustrated in Expression \ref{eq:queryAnswerRegister}.

\begin{equation}
\ket{q} \ket{a} = \frac{1}{\sqrt{2^{n}}}\sum_{x = 0}^{2^{n} - 1} \ket{ x } \ket{0}
\label{eq:queryAnswerRegister}
\end{equation}

Additionally, suppose that a quantum oracle with the form presented in Expression \ref{eq:oracleDefinition} is constructed. The auxiliary function $\phi: \{0,1\}^{n} \rightarrow \{0,1\}$ employed simply verifies if an argument is a solution or not for a specific SAT instance, as illustrated by Expression \ref{eq:oracleFunctionDefinition}. We should be careful to point out that an efficient oracle responsible for verifying the validity of a variable configuration for a specific $\phi$ can be easily constructed by mapping the $m$ boolean connectives of the network onto a reversible circuit (see \cite{bennett1973}) in order to ensure a unitary mapping.

\begin{equation}
O \ket{q} \ket{a} = \ket{q} \ket{a \oplus \phi(q)}
\label{eq:oracleDefinition}
\end{equation}

\begin{equation}
\phi(q) = \phi(x_{1}, x_{2}, \cdots, x_{n}) = \left\{
		\begin{array}{ll}
		1 & \text{if $q$ evaluates to true}\\
		0 & \text{otherwise}
		\end{array} \right.
\label{eq:oracleFunctionDefinition}
\end{equation}

If oracle $O$ is applied to the combined state of Expression \ref{eq:queryAnswerRegister} a result like the one illustrated in Expression \ref{eq:queryRegisterAnswerAfterOracle} may be obtained, where $\ket{\psi^{\prime}}$ denotes the overall superposition evaluation. For simplification issues we assume that there exists at least a solution. Naturally, some of the query values produce a solution, whilst others do not. 

\begin{equation}
\ket{\psi^{\prime}} = \frac{1}{\sqrt{2^{n}}}\sum_{x = 0}^{2^{n} - 1} O \ket{ x } \ket{0} = 
	\begin{cases} 
	\ket{\underbrace{00 \cdots 0}_{\text{n bits}} } \ket{0}\\
	\ket{00 \cdots 1 } \ket{0}\\
	\quad \quad \vdots\\
	\ket{11 \cdots 0 } \ket{1}\\
	\ket{11 \cdots 1 } \ket{0}
	\end{cases}
\label{eq:queryRegisterAnswerAfterOracle}
\end{equation}

From this point on the system's state can no longer be expressed as a tensor product between query and answer register, i.e. the system becomes entangled. Quantum entanglement is a key feature of quantum mechanics which details the connections between subsystems of compound quantum systems. It was a key aspect of the quantum world formalism proposed by von Neumann in 1932 \cite{neumann1932}. Although the intriguing impacts of quantum inseparability were only later grasped by Einstein, Podolsky and Rosen \cite{einstein1935} alongside Schr{\"o}dinger \cite{schrodinger1935}. Quantum entanglement is also a key resource in quantum information.

Mathematically, we can describe the state of each register by tracing out the remaining register, through the partial trace mechanism. In this case we are interested in the overall state of the answer register. In order to calculate the partial trace of the answer register we first need to calculate $\varrho$ the density operator of the quantum state presented in Expression \ref{eq:queryRegisterAnswerAfterOracle}. The overall form for $\varrho^{a}$ is illustrated in Expression \ref{eq:partialTraceOfAnswerRegister}

\begin{equation}
\begin{array}{lll}
\varrho &=& \ket{\psi} \bra{\psi} \\
	    &=&  \frac{1}{\sqrt{2^{n}}}( \ket{00 \cdots 0 } \ket{0} + \ket{00 \cdots 1 } \ket{0} + \cdots + \ket{11 \cdots 0 } \ket{0} + \ket{11 \cdots 1 } \ket{0})\\
	    & & \frac{1}{\sqrt{2^{n}}}( \bra{00 \cdots 0 } \bra{0} + \bra{00 \cdots 1 } \bra{0} + \cdots + \bra{11 \cdots 0 } \bra{0} + \bra{11 \cdots 1 } \bra{0})\\
	    &=& \frac{1}{2^{n}} \ket{00 \cdots 0} \ket{0} ( \bra{00 \cdots 0 } \bra{0} + \bra{00 \cdots 1 } \bra{0} + \cdots + \bra{11 \cdots 0 } \bra{0} + \bra{11 \cdots 1 } \bra{0}) + \\ 
	     & & \frac{1}{2^{n}} \ket{00 \cdots 1} \ket{0} ( \bra{00 \cdots 0 } \bra{0} + \bra{00 \cdots 1 } \bra{0} + \cdots + \bra{11 \cdots 0 } \bra{0} + \bra{11 \cdots 1 } \bra{0}) + \\
	    & & + \cdots + \\
	    & & \frac{1}{2^{n}} \ket{11 \cdots 0} \ket{1} ( \bra{00 \cdots 0 } \bra{0} + \bra{00 \cdots 1 } \bra{0} + \cdots + \bra{11 \cdots 0 } \bra{0} + \bra{11 \cdots 1 } \bra{0}) + \\
	    & & \frac{1}{2^{n}} \ket{11 \cdots 1} \ket{0} ( \bra{00 \cdots 0 } \bra{0} + \bra{00 \cdots 1 } \bra{0} + \cdots + \bra{11 \cdots 0 } \bra{0} + \bra{11 \cdots 1 } \bra{0}) + \\
\end{array}
\label{eq:overrallCompositeEntangledState}
\end{equation}

\begin{equation}
\begin{array}{lll}
\varrho^{a} 	&=& \text{Tr}_{q}( \varrho )\\
			&=& \frac{1}{2^{n}} ( \bra{00 \cdots 0} \ket{00 \cdots 0} \ket{0}\bra{0} + \bra{00 \cdots 1}\ket{00 \cdots 1} \ket{0}\bra{0} +\\
			& & + \cdots + \\
			& & \bra{11 \cdots 0}\ket{11 \cdots 0} \ket{1}\bra{1} + \bra{11 \cdots 1}\ket{11 \cdots 1} \ket{0}\bra{0})\\
			&=& \frac{1}{2^{n}} \left[ (2^{n} - 1) \ket{0}\bra{0} + \ket{1	}\bra{1}\right]
\end{array}
\label{eq:partialTraceOfAnswerRegister}
\end{equation}

Generally, the result presented in Expression \ref{eq:overrallCompositeEntangledState} can be improved if we take into account the number of solutions. Accordingly, let $k$ denote the overall number of solutions, then $\varrho^{a}$ takes the form shown in Expression \ref{eq:partialTraceOfAnswerRegisterWithVariableForSolutions}. Notice that the overall state is separable only when $k = 0$, i.e. no solution exists, or when $k = 2^{n}$, each value belonging to $[0, 2^{n} - 1]$ is a solution. Otherwise, the system is entangled.

\begin{equation}
\varrho^{a} = \frac{1}{2^{n}} \left[ (2^{n} - k) \ket{0}\bra{0} + k\ket{1}\bra{1}\right]
\label{eq:partialTraceOfAnswerRegisterWithVariableForSolutions}
\end{equation}

Thus, the problem of determining whether or not a solution to a problem exists can be reduced to the problem of determining whether the overall quantum state is separable or entangled. 

\subsection{Quantum entanglement detection \label{sec:quantumEntanglementDetection}}

The quantum separability problem consists in determining if a given a density matrix $\varrho$ representing a quantum state is entangled or separable \cite{gharibian2008}. Efficiently deciding on the nature of such states has grabbed researchers attention and remains a problem of crucial importance to the fields of quantum computation and information \cite{ioannou2006b}. Generally speaking, quantum entanglement is studied in accordance with a varied mix of properties (just to name a few of these: bipartite vs. multipartite systems, pure vs. mixed states, bound entanglement; for exhaustive reviews please refer to \cite{krammer2005}, \cite{horodecki2007} and \cite{guhne2009}). It is important to mention that the quantum separability question has been approached from the classical and quantum perspectives. These approaches typically consider the nature of the input (classical vs. quantum), and whether any required processing will be performed on a classical or quantum computer \cite{ioannou2006a}. This problem was shown to be NP-hard classically \cite{gurvits2003}. However, as mentioned in \cite{ioannou2006a} the processes involving both quantum input and processing have not been thoroughly investigated.  

In the case of our specific approach we would only need to consider bipartite quantum systems with mixed states. As pointed out in \cite{horodecki1996} the mixed state requirement stems from the fact that any potential laboratory demonstration of this approach would have to deal with mixed states rather than pure ones, due to the uncontrolled interactions with the environment. These requirements are present in one of the existing quantum detection schemes, namely the one proposed in \cite{horodecki2002}. The method employed by the authors is experimentally viable and provides for a direct detection mechanism of quantum entanglement. Their approach is based on the theoretical foundations laid down in \cite{horodecki1996}. The method determines whether a state $\varrho$ is separable or not, i.e. entangled, based on the mathematical properties of linear positive maps acting on matrices. More specifically \cite{horodecki2007}, let $M_{d} \rightarrow M_{d}$ be the space of matrices of dimension $d$, a map $\Lambda: M_{d} \rightarrow M_{d}$ is called positive if it is Hermitian and has non-negative spectrum. Additionally, the map $\Lambda$ is completely positive if and only if $I \otimes \Lambda$ is positive for identity map $I$ on any finite-dimensional system. A state $\varrho$ is separable if and only if the result presented in Expression \ref{eq:separabilityCriterion} is observed for all positive but not completely positive maps $\Lambda: M_{d} \rightarrow M_{d}$.

\begin{equation}
[I \otimes \Lambda](\varrho) \geq 0
\label{eq:separabilityCriterion}
\end{equation}

%Expression \ref{eq:separabilityCriterion} implies that, despite its form, map $\Lambda$ always exists. 

Expression \ref{eq:separabilityCriterion} cannot be directly used since it requires knowing state $\varrho$ beforehand. Additionally, positive maps $\Lambda$ cannot be directly implemented in laboratory. Fortunately, it is possible to obtain a physically realizable map by mixing an appropriate proportion of $[I \otimes \Lambda]$ with a depolarizing map. This approach allows for a new map $[ \widetilde{I \otimes \Lambda}]$ to be obtained, which have been referred to as structural physical approximations. For more on this subject please refer to \cite{fiurasek2002}. The separability criterion can then be restated as follows \cite{horodecki2002}: $\varrho$ is separable if and only if for all positive maps $\Lambda$ the condition presented in Expression \ref{eq:separabilityCriterionMixedMapEigenvalue} is observed.

\begin{equation}
[\widetilde{I \otimes \Lambda}](\varrho) \geq \frac{d^{2} \lambda}{d^{4}\lambda + 1}
\label{eq:separabilityCriterionMixedMapEigenvalue}
\end{equation}

Where $\lambda$ corresponds to the most negative eigenvalue obtained when the induced map $[(I \otimes I) \otimes (I \otimes \Lambda)]$ acts on the maximally entangled state of the form $\frac{1}{d^{2}}\sum_{i=1}^{d^{2}}\ket{i}\ket{i}$. Accordingly, Expression \ref{eq:separabilityCriterionMixedMapEigenvalue} states that the lowest eigenvalue of the transformed state $\varrho^{\prime} = [\widetilde{I \otimes \Lambda}](\varrho)$ should be greater than $\frac{d^{2} \lambda}{d^{4}\lambda + 1}$ for $\varrho$ to be separable. 

The authors devised a method which allows for an estimate of the lowest eigenvalue to be obtained efficiently and directly. It requires that a joint measurement be performed on $N$ copies of state $\varrho^{\prime}$. The overall input density operator of the estimation problem is $\varrho^{\prime \otimes N}$, which exists on the $N$th tensor power $\mathcal{H}^{\otimes N}$ \cite{keyl2001}. The error $\epsilon$ associated with the estimate of the lowest eigenvalue decreases exponentially with $N$. Such a measurement can be represented as quantum network implementing projections on the symmetric and partially symmetric subspaces \cite{horodecki2002}. An efficient method addressing these questions was proposed in \cite{barenco1997} requiring a number of auxiliary gates that grows quadratically with the dimension of the input, i.e. $O(n^{2})$, where $n$ is the number of bits. If $\varrho^{\prime}$ represents the state of an $n$ qubit register, then each additional tensor power will mean that another $n$ bits should be accounted. Consequently, an $\varrho^{\prime \otimes N}$ system will have a total of $N \times n$ bits. Which means that the quantum network responsible for estimating the lowest eigenvalue will have $O(N^{2}n^{2})$ complexity. 

Clearly, this approach is dependent on map $\Lambda$ which have not been operationally characterized so far \cite{michalHorodecki2001}. As pointed out in \cite{horodecki2007} in general the set of positive but not completely positive maps is not characterized and it involves a hard problem in contemporary linear algebra. However, for low dimensional systems, namely those with dimension $2 \otimes 2$ or $2 \otimes 3$, the positive partial transpose map proposed in \cite{peres1996} can be employed as the $\Lambda$. In \cite{michalHorodecki2001} the authors draw attention to the fact that `Recently, the progress in this direction has been made \cite{lewenstein2000} \cite{lewenstein2001} which suggests that tests of separability based on positive maps will soon acquire practical meaning beyond the scope of two-qubit systems.'' Whether such a map $\Lambda$ acting on $\mathcal{H}_{d} \otimes \mathcal{H}_{d}$ quantum systems can be determined remains an open question.

% SEE LAST PARAGRAPH PAGE 8/17 OF \cite{BARENCO} IN ORDER TO UNDERSTAND THAT THE SIZE OF THE NETWORK GROWS QUADRATICALLY WITH THE DIMENSION OF THE INPUT.

\subsection{Subset entanglement inducing oracle \label{sec:subsetEntanglementInducingOracle}}

The $\Omega(\sqrt{N})$ lower bound for quantum search employing oracles working on the full range of searchable items implies that an alternative search approach has to be devised. In classical tree search it is a standard technique to start by analyzing subtrees  and deciding whether these may eventually lead to a solution. Based on problem requirements it is possible to automatically exclude, i.e. prune, certain subtrees. The act of pruning may eventually be responsible for large sections of the tree to be discarded, and therefore allow the search to terminate faster. We will draw inspiration from these concepts of classical search in order to develop our approach to quantum hierarchical search. 

Quantum algorithms employing traditional oracles provide at most a polynomial advantage over classical algorithms for total functions, i.e. functions defined for the whole of $\{0,1\}^{n}$, where $n$ is the number of bits. The oracle model contemplates syperpolynomial advantage but only when partial functions are defined which operate on a subset of $\{0,1\}^{n}$ \cite{kaye2007a}. Notice that classical search can be viewed as a procedure which evaluates subsets of an initial range. Since in quantum computation the oracle operator can be applied to a superposition of computational basis, evaluating subsets is equivalent to only evaluating specific ranges of the superposition. Accordingly, it is possible to develop an oracle responsible for evaluating only a certain subset of the initial range $[0, 2^{n} - 1]$ allowed with $n$ qubits. Although we are only interested in evaluating specific subset there are other alternatives for trying to decompose a quantum search space. For instance, Grover concluded in \cite{grover2004a} that determining the first $n$ bits of a solution by employing amplitude amplification schemes is only slightly easier than determining the total bits.

This model for a range specific entanglement inducing oracle can be described as presented in Expression \ref{eq:unitaryOperatorWithRange} which employs an auxiliary function $f_{[a,b]}(q)$ defined in Expression \ref{eq:oracleFunctionWithRange}. In the case of the SAT problem it would be convenient to define $f_{[a,b]}(q)$ as $\phi_{[a,b]}(q)$.

\begin{equation}
	O_{[a,b]} \ket{ q } \ket{a}  = \ket{q} \ket{ a \oplus f_{[a,b]}(q)}
\label{eq:unitaryOperatorWithRange}
\end{equation}

\begin{equation}
	f_{[a,b]}(q) = 
	\begin{cases}
	1 & \text{if } f(q) \text{ is a solution and } q \in [a,b] \\
	0 & \text{otherwise}
	\end{cases}
\label{eq:oracleFunctionWithRange}
\end{equation}

As was previously pointed out, the oracle evaluation process has the overall effect of making the quantum registers entangled. By testing whether the oracle has induced, or not, quantum entanglement it is possible to decide if the search procedure should continue decomposing a range, or if another range should be tested. Ideally, the entanglement detection scheme should present some type of polynomial upper-bound behavior such as the one described in the previous section. \footnote{The entanglement detection approach described in \cite{horodecki2002} requires the overall bipartite system to be $d \otimes d$. Consequently, the answer register $\ket{a}$ should have the same dimension than $\ket{q}$, i.e. $n$ bits. This requirement has no direct consequences in the overall oracle since unitary evolution can still be assured.} If the state $\varrho$ resulting from applying an oracle $O$ with the form presented in Expression \ref{eq:oracleFunctionWithRange} is separable then the range evaluated can automatically be discarded. Discarding a wide range of potential candidates \textit{en masse} can be understood as the classical tree search operation of pruning certain subtrees.  On the other hand, if $\varrho$ is entangled then it is possible to further decompose the associated range. Eventually, this sort of recursive branch and bound procedure, by constantly readjusting the range of oracle $O$, will ``zoom in'' on a solution. Extending this method to problems presenting multiple solutions would require systematically focusing on previously non-expanded but solution-containing ranges. 

Notice that this approach requires  a new oracle to be defined with each iteration in terms of a specific subset that may be entangled. The set of oracles applied throughout the search can be viewed as a single ``dynamic'' oracle, which differs substantially from the standard ``static'' oracle applied in quantum search. Additionally, in contrast with Grover's algorithm, we are not interested in performing an amplitude amplification process, but rather we are concerned with decomposing the quantum search space.

\subsection{On the growth of the number of copies required  \label{ch:quantumEntanglementAndPartialSearch:sec:onTheGrowthOfTheNumberOfCopiesRequired}}

Clearly, one question still lingers: What can be said about the number of copies $N$ of the system that are required? According to \cite{kaye2007a} any procedure that on input $\ket{\psi_{Z}}$ guesses whether Z = X or Z = Y will guess correctly with probbability at most $1 - \epsilon = \frac{1}{2} + \frac{1}{2}\sqrt{1-\delta^{2}}$ where $\delta = | \bra{\psi_{x}} \ket{\psi_{y}}|$. For our particular case we are interested in distinguishing two specific cases, namely:

\begin{itemize}

	\item $\ket{\psi_{solution}}$ which results from applying $U\ket{\psi}$ when \textit{one} solution exists;

	\item $\ket{\psi_{no-solution}}$  which results from applying $U\ket{\psi}$ when \textit{no} solution exists;
	
\end{itemize}

For search spaces of dimension $L$ the initial amplitudes $\alpha_{i}$ associated with each computational basis $i$ of the superposition $\ket{\psi}$ is $\frac{1}{\sqrt{L}}$. After having applied oracle $O$ the two states remain exactly equal except for two computational basis where the amplitudes permuted. This means that when calculating the inner product the permuted computational basis will sum up to zero. Accordingly, the inner product will sum the value $\frac{1}{\sqrt{L}}^{2}$ a total of $L-1$ times, i.e.

\begin{equation}
\delta = \bra{\psi_{solution}}\ket{\psi_{no-solution}} = \frac{L-1}{L}
\label{ch:quantumEntanglementAndPartialSearch:eq:innerProduct}
\end{equation}

\begin{figure}[ht]
	\centering
	\includegraphics[width=.9\textwidth]{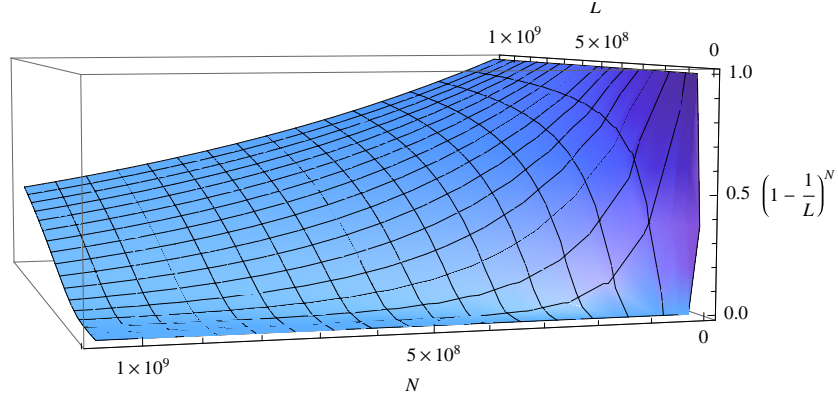}
	\caption{Three-dimensional plot of $\delta^{\otimes N}$ as a function of $L \in [2^{1},2^{30}]$ and $N \in [2^{1},2^{30}]$. \label{ch:quantumEntanglementAndPartialSearch:fig:growthInTheNumberOfCopies}}
\end{figure}

Given a tensor product of $N$ items, Expression \ref{ch:quantumEntanglementAndPartialSearch:eq:innerProduct} evolves into Expression \ref{ch:quantumEntanglementAndPartialSearch:eq:innerProductNtimes}. The three-dimensional plof of $\delta^{\otimes N}$ as a function of $L \in [2^{1},2^{30}]$ and $N \in [2^{1},2^{30}]$ is illustrated in Figure \ref{ch:quantumEntanglementAndPartialSearch:fig:growthInTheNumberOfCopies}.

\begin{equation}
\delta^{\otimes N } = \bra{\psi_{solution}}\ket{\psi_{no-solution}}^{\otimes N} = \frac{L-1}{L}^{N}
\label{ch:quantumEntanglementAndPartialSearch:eq:innerProductNtimes}
\end{equation}

In order for these states to be distinguished with significant probability the inner product $\delta^{\otimes N}$ presented in Expression \ref{ch:quantumEntanglementAndPartialSearch:eq:innerProductNtimes} must be made small. However, in order to achieve this one needs to chose a number of copies $N$ that grows in accordance with the dimension of the search space $L$, i.e. $N = O(L)$. Consequently, this approach would not provide for any gains over classical search.

\subsection{Consequences for efficient entanglement detection schemes \label{sec:consequencesForEfficientEntanglementDetectionSchemes}}

What would be the consequences if the number of system copies $N$ was not a function of the search space? Suppose the proposed search procedure is executed on $n$-qubits placed on a superposition. Initially, the algorithm has to decompose the $[0, 2^{n} - 1]$ initial range. Lets assume that any specific range being considered is split in half. Accordingly, the procedure needs to verify if evaluating the elements in $[0, 2^{n - 1} - 1]$ produces an entangled quantum state $\varrho$. If this is found to be true then subset  $[0, 2^{n - 1} - 1]$ can be also split in half and evaluated. Otherwise, subset $[2^{n - 1} - 1, 2^{n} - 1]$ needs to be decomposed. Independently of what subset induces entanglement, the algorithm is able to prune half of the $2^{n}$ initial states, i.e. $2^{n}/2$. Accordingly, for iteration $i$, the oracle is able to focus on $2^{n}/2^{i}$ states. Clearly, when $i = n$ a single state is being considered and consequently a solution can be determined with certainty by employing $O(n)$ oracle queries. Associated with each oracle query is the quantum entanglement detection scheme bringing the overall complexity of our approach to $O(N^{2}n^{3})$. In the case of the SAT problem we have to consider the costs associated with each oracle query, respectively $n + m$. Consequently, a solution for SAT would be calculated in $O(N^{2}n^{4} + N^{2}n^{3}m)$ quantum polynomial time. 

It is our believe that it is not possible to efficiently detect quantum entanglement. If we take into account the simplicity of the search procedure designed in Section \ref{sec:subsetEntanglementInducingOracle} then if such a method existed we could efficiently search, i.e. in quantum polynomial time, a problem space of dimension d. Accordingly, we can define the following conjecture.

\begin{description}

	\item[Quantum entanglement detection conjecture] - It is not possible to efficiently detect quantum entanglement non-classically since this would automatically imply that a simple algorithm exists proving that NP=EQP.
	
\end{description}

The above conjecture stresses the notion that there appears to be a relationship between entanglement detection and search in terms of computational complexity.
 
\section{Conclusions \label{sec:conclusions}}

The general characterization of positive but not completely positive linear maps $\Lambda$ alongside quantum entanglement detection schemes and partial range entanglement inducing operators may eventually be responsible for producing efficient algorithmic solutions capable of searching exponential-growth search spaces. Although some research has already been carried out, further thorough analysis into the subject is still required. However, given that $N = O(L)$ current methods cannot be employed in order to speed up quantum search.

%TODO
%
%\begin{itemize}
%	
%	\item We are basically saying that there is a relationship between entanglement detection and search in terms of complexity.
%\end{itemize}

%%%%%%%%%%%%%%%%%%%%%%%%%%%%%%%%%%%%%%%%%%%%%%%%%%%%%%%%%%%%
%       									BIBLIOGRAPHY					         		     %
%%%%%%%%%%%%%%%%%%%%%%%%%%%%%%%%%%%%%%%%%%%%%%%%%%%%%%%%%%%%
%\bibliographystyle{ieeetr}
\bibliographystyle{splncs}
%\bibliography{../../../../../Bibliography/bibliography}

\end{document}